\documentclass[aps,prl,twocolumn,amsmath,showpacs,superscriptaddress]{revtex4}
\usepackage{bm}
\usepackage{amsmath}
\usepackage{amssymb}
\usepackage{amsfonts}
\usepackage{graphicx}
\usepackage{color}

\begin{document}

\title{Screening of Coulomb Impurities in Graphene}

\author{Ivan S. Terekhov}
\affiliation{School of Physics, University of New South Wales,
Sydney 2052, Australia}
\affiliation{Budker Institute of
Nuclear Physics, 630090 Novosibirsk, Russia}

\author{Alexander I. Milstein}
\affiliation{Budker Institute of
Nuclear Physics, 630090 Novosibirsk, Russia}

\author{Valeri N. Kotov}
\affiliation{Department of Physics,
Boston University, 590 Commonwealth Avenue, Boston, Massachusetts 02215}

\author{Oleg P. Sushkov}
\affiliation{School of Physics,
University of New South Wales, Sydney 2052, Australia}


\begin{abstract}
We calculate exactly  the vacuum polarization charge density in the field of a
 subcritical Coulomb impurity, $Z|e|/r$, in graphene. Our analysis is based on the
exact electron Green's function,  obtained  by using the operator method, and
leads to  results that are exact in the parameter $Z\alpha$, where $\alpha$ is the
``fine structure constant" of graphene. Taking into account also
electron-electron interactions in the Hartree approximation, we
solve the problem self-consistently in the subcritical regime,
where the impurity has an effective charge $Z_{\mbox{eff}}$, determined by the
localized induced charge.
 We find that an impurity with bare charge $Z=1$ remains subcritical,
$Z_{\mbox{eff}} \alpha < 1/2$, for any  $\alpha$, while impurities with
$Z=2,3$ and higher can become supercritical at certain values of $\alpha$.
\end{abstract}


\maketitle

It has been known for a long time that the single electron
dynamics in a monolayer of graphite (graphene) is described by a
massless two-component Dirac equation
\cite{Wallece,McClure,Gonzalez}. A surge of interest in the
problem was caused by the recent successful fabrication of
graphene \cite{Novoselov1} and measurements of transport
properties \cite{Novoselov05,Zhang,Novoselov06,Morozov06,Zhang06,Chen07},
including an unconventional form of the quantum Hall effect.
Due to the Coulomb interaction between electrons, graphene
represents a peculiar two-dimensional (2D) version of massless
Quantum Electrodynamics (QED) \cite{Gonzalez}. It appears to be
much simpler than conventional QED because the interaction is
described by the instantaneous $1/r$ Coulomb's law. On the other
hand the Fermi velocity  $v_F \approx 10^6 \mbox{m/s} \approx
c/300$ ($c$ is the velocity of light), and therefore the
 ``fine structure constant''  $\alpha=e^2/\hbar v_F
\sim1$, leading to a  strong-coupling version of QED. Below we set
$\hbar=v_F=1$. Screening of a charged nucleus due to vacuum
polarization is an effect of fundamental importance in QED. This
problem was investigated in detail both in the subcritical and
supercritical regimes \cite{WichmannKroll,McLerran,Mil2,Zeld}.
The problem of charged impurity screening in graphene, which also can be treated
in terms of vacuum polarization, has recently received a lot of attention
\cite{DiVincenzo,nomura,Ando,Hwang,Katsnelson,Shytov,pereira,Subir,Fogler},
due to the importance of the problem for transport properties involving
 charged impurities, as well as for our general understanding  of  the theory of graphene.

To leading order in the weak coupling expansion, $Z\alpha \ll 1$,
the induced charge is negative and  localized at the impurity
position, $\rho_{ind}=-|e|\frac{\pi}{2}(Z\alpha)\delta(\mathbf{r})$, which
leads to screening of the impurity potential
\cite{Kolezhuk,Shytov,pereira,Subir}. We denote by $Z|e|$ the impurity charge,
 and $e=-|e|$ is the effective electron charge; from now on we refer to
 $Z$ as the impurity charge with the understanding that it is measured in units
 of $|e|$.
In graphene, the  strong-coupling  problem $Z\alpha \sim 1$ was recently addressed \cite{Shytov},
 and it was found that the supercritical regime occurs for
 $Z\alpha > 1/2$, where a  $1/r^2$ tail appears in  the induced charge density,
 while in the subcritical regime  $Z\alpha < 1/2$, the induced charge is  always
localized at the impurity site.
Analytical results were also supplemented by numerical lattice
 calculations \cite{pereira}, leading to similar conclusions.
The induced charge behavior for small $Z\alpha$    was  first emphasized
   in a different context in Ref.~\cite{Kolezhuk}, and
 it is in  agreement with a recent perturbative calculation of non-linear
vacuum polarization at order $(Z\alpha)^{3}$ \cite{Subir}.  It
   was also pointed out \cite{Subir}  that a power law tail can appear even
 in the subcritical regime due to interaction effects
(at order  $Z\alpha^{2}$).

In the present paper we investigate analytically the induced
charge density in graphene in  the subcritical regime. We use the
method of calculation suggested in Ref.~\cite{Mil2}, where the
the induced charge density in a strong Coulomb
field was obtained in coordinate space in three-dimensional (3D) QED.
We express the induced charge density via the exact  Green's function
in a Coulomb field, calculated within  the operator technique \cite{MS82}.
Our main result is an exact expression for  the polarization charge,
 non-perturbative in the parameter $Z\alpha$, and we explore its
physical consequences. The exact result allows us to determine
the effective impurity charge $Z_{\mbox{eff}}$ in a self-consistent way.
Perhaps most surprisingly, we find that an impurity with bare
 charge $Z=1$ can never become supercritical, i.e.  $Z_{\mbox{eff}}\alpha < 1/2$.
In spite of this, screening can be very substantial, i.e. one can have
$Z_{\mbox{eff}} \ll 1$ for large enough $\alpha$.

{\it Electron Green's function in a Coulomb field}.
It is convenient, for technical reasons,
 to introduce a small ``electron mass'' $M$  which
 we set to zero at the end of our calculations. This will allow
 us to avoid some difficulties which appear in the calculation
of the induced charge in massless QED \cite{McLerran}, and will
 serve the renormalization purpose. Physically the mass
describes a small energy splitting between carbon atoms in the unit cell.
We will consider  only the  half-filled case of graphene, i.e.
we  set  the chemical potential to zero.

The  electron  Green's function $G(\mathbf{r},\mathbf{r}'|\epsilon)$
in a Coulomb field satisfies the two-component equation
\begin{equation}
\label{GFE}
\left(\epsilon +\frac{Z\alpha}{r} -
({\bm{\sigma}}\cdot\mathbf{p}) - \sigma_3
M\right)G(\mathbf{r},\mathbf{r}'|\epsilon)=\delta(\mathbf{r}-\mathbf{r}').
\end{equation}
 Here ${\bm{\sigma}}=(\sigma_1,\sigma_2)$, and $\sigma_{1,2,3}$ are the
Pauli matrices; $\mathbf{p}=(p_x,p_y)$ is the momentum operator.
Following Ref.~\cite{MS82} we represent the solution of
Eq.~(\ref{GFE}) in the form
\begin{eqnarray}
\label{GreenOperForm3}
&&G(\mathbf{r},\mathbf{r}'|\epsilon)=-i\left(\epsilon
+\frac{Z\alpha}{r} +
({\bm{\sigma}}\cdot\mathbf{p}) + \sigma_3 M\right) \nonumber \\
&&\times\int_0^\infty ds \,e^{2is\epsilon Z\alpha}
\exp\left(is[r\Delta_r+\frac{\hat{K}}{r}+r(\epsilon^2-M^2)]
\right) \nonumber\\
&& \times \sqrt{\frac{r}{r'}}\delta(r-r')\delta(\phi-\phi') ,
\end{eqnarray}
where $\Delta_r$ is the radial part of the Laplace operator and
$\hat{K}=\partial^2/\partial\phi^2+(Z\alpha)^2-iZ\alpha({\bm{\sigma}}\cdot\mathbf{n})$,
 ${\bf n}={\bf r}/r$.
Then we express $\delta(\phi-\phi')$ via the projectors $P_\lambda$
that are eigenfunctions of the operator $\hat{K}$, $\hat{K}P_\lambda=-\lambda^2 P_\lambda$:
\begin{eqnarray}
\label{ProjectionOperators}
&&\delta(\phi-\phi')= \sum_\lambda
P_\lambda(\phi,\phi')\, , \\
&& \lambda = \gamma\mp\frac{1}{2}\ ,  \  \gamma =
\sqrt{\varkappa^2-(Z\alpha)^2}\ ,  \ \varkappa = m+1/2 \ ,
\nonumber
\end{eqnarray}
and $m=0,1,2, ...$. The explicit form of $P_\lambda$ for  $\lambda
= \gamma - \frac{1}{2}$ is
\begin{eqnarray}
\label{ProjectionOperators1} &&
P_\lambda(\phi,\phi')=\frac{1}{4\pi\gamma}
\begin{pmatrix}
P_{11}&P_{12}\\
P_{21}&P_{22}
\end{pmatrix},\\
&&  P_{11} = (\gamma+\varkappa)e^{im(\phi-\phi')}
+(\gamma-\varkappa)e^{-i(m+1)(\phi-\phi')}, \nonumber\\
&&  P_{12} = -iZ\alpha\left(e^{-i(m+1)\phi}e^{im\phi'} +
e^{im\phi}e^{-i(m+1)\phi'} \right), \nonumber\\
&& P_{11} = P_{22}^{\ast}, \ P_{12} = -P_{21}^{\ast}. \nonumber
\end{eqnarray}
The projector with eigenvalue $\lambda = \gamma+ 1/2$ is obtained
from (\ref{ProjectionOperators1}) by  replacing
 $\gamma\to-\gamma$.
After substitution of the expansion (\ref{ProjectionOperators})
into Eq.~(\ref{GreenOperForm3}), the problem is reduced to
calculation of the action of the operator
$\exp\left(-2is\left[A_1+k^2 \frac{A_3}{2} \right] \right)$ on
$\sqrt{\frac{r}{r'}}\delta(r-r')$. Here $k^2 =M^2-\epsilon^2$ and
the operators are defined as:
$A_1=\frac{1}{2}\left(-\frac{\partial}{\partial
r}r\frac{\partial}{\partial r} +\frac{\lambda^2}{r}\right),
A_3=r$. Along with $A_2=-i(r\partial/\partial r+1/2)$, these
operators generate an $O(2,1)$ algebra, which was considered in
Ref.~\cite{MS82} in relation to the 3D Green's function for the  Dirac
equation of an electron in a Coulomb field. Therefore we can
directly use the operator transformation of that work. As a
result we find the following integral representation for the
solution of Eq.~(\ref{GFE}):
\begin{eqnarray}\label{GreenFinal}
&&G(\mathbf{r},\mathbf{r}'|\epsilon)=-\left(\epsilon
+\frac{Z\alpha}{r} +
({\bm{\sigma}}\cdot\mathbf{p}) + \sigma_3 M\right) \\
\nonumber
&&\times\sum_{\lambda}P_{\lambda}(\phi,\phi')\int_0^\infty ds \,
\frac{k}{\sin(ks)}e^{2is\epsilon Z\alpha}\nonumber\\
&&\times\exp\left[ i
k(r+r')\cot(ks)-i\pi\lambda\right]J_{2\lambda}\left(\frac{2k\sqrt{r
r'}}{\sin(ks)}\right)\, .\nonumber
\end{eqnarray}
Here $J_{\nu}(x)$ is the Bessel function.

{\it Induced charge}. The induced charge density is
\begin{eqnarray}\label{InducedGhargeGeneral}
\rho_{ind}(\mathbf{r})=-ieN\int_{C}\frac{d\epsilon}{2\pi}\mathrm{Tr}\{
G(\mathbf{r},\mathbf{r}|\epsilon)\}\, ,
\end{eqnarray}
where $G$ is the Green's function calculated above,
and  $N=4$ reflects the spin and valley degeneracies. The contour
of integration $C$ goes below the real axis in the left half-plane
and above the real axis in the right half-plane of the complex
$\epsilon$ plane.
Taking into account the analytical properties of the Green's
function, the contour of integration with respect to $\epsilon$
 can be deformed to coincide with the imaginary axis. The integration
contour with respect to $s$ in  Eq.~(\ref{GreenFinal}) can then
also be rotated to coincide with the imaginary axis so that it extends
from zero to $-i\infty$ for $\mathrm{Im}\epsilon>0$, and from zero
to $i\infty$ for $\mathrm{Im}\epsilon < 0$.  After these
transformations and obvious change of variables, we obtain:
\begin{eqnarray}
\label{InducedGhargeR}
&&\rho_{ind}(r)=-N\frac{e}{\pi^2
r}\sum_{m=0}^{\infty}\int\limits_{0}^{\infty}\int\limits_{0}^{\infty}
 d\epsilon\, ds\,e^{-y\cosh s} \\
&&\times\left( 2Z\alpha \cos(\mu s)\coth s I_{2\gamma}(y)-\sin(\mu
s)\frac{\epsilon}{k} y I'_{2\gamma}(y)\right)\, ,\nonumber
\end{eqnarray}
where $k=\sqrt{\epsilon^2+M^2}\,$, $y=2rk/\sinh s$,
$\mu=2Z\alpha\epsilon/k\,$, $I_{\gamma}(x)$ is the modified Bessel
function of the first kind, and $I'_{2\gamma}=dI_{2\gamma}(y)/dy$.
We note that $\rho_{ind}(r)$, Eq.~(\ref{InducedGhargeR}), is an
odd function of $Z\alpha$.

In fact, the expression (\ref{InducedGhargeR}) is not well defined and the answer depends
on the order of integration over $\epsilon$ and $s$. To overcome
this problem we follow the usual procedure of QED. We perform the
regularization of the integrals introducing a finite upper limit
of integration for $\epsilon$, and a lower limit of integration for
$s$. Then we carry out the renormalization using the obvious
physical requirement of zero total induced charge. We can satisfy
this requirement due to the non-zero mass $M$ because  the induced charge
density diminishes rapidly at distances $r\gg 1/M$. After
renormalization the result is independent of the cutoff parameters
and the order of integration. It is technically convenient to do the
renormalization in  momentum space. All details of calculations
which we need to perform  are similar to those  described
in detail in Ref.~\cite{Mil2} for the problem of 3D vacuum
polarization in QED. Finally, we obtain the renormalized induced
charge density in momentum representation $\rho_{ind}^{R}(q/M)$.
The leading term of the asymptotics of this function at $M\to 0$ (or
$q/M\to\infty$) is a constant,  $Q_{ind}$. Therefore the induced
charge density in coordinate space has the form
\begin{equation}
\label{rind}
\rho_{ind}(\mathbf{r})=Q_{ind} \delta(\mathbf{r})+\rho_{distr}\ .
\end{equation}
For the induced charge $Q_{ind}$  we find
\begin{eqnarray}
\label{QinQRepresentation}
&&Q_{ind}=eN\left[\frac{\pi}{8}Z\alpha+
\Lambda(Z\alpha)\right]\equiv -|e|A(Z\alpha)\,, \nonumber\\
&&\Lambda(Z\alpha)=\frac{2}{\pi}\sum_{m=0}^{\infty}\mathrm{Im}\biggl[\ln\Gamma(\gamma-iZ\alpha)
+\frac{1}{2}\ln(\gamma-iZ\alpha)\biggr. \nonumber\\
&&-\left.(\gamma-iZ\alpha)\psi(\gamma-iZ\alpha)+\frac{iZ\alpha}{2
\varkappa} -iZ\alpha \varkappa  \psi'(\varkappa)\right] \ ,
\end{eqnarray}
where $\Gamma(x)$ is the gamma function and  $\psi(x)=d\ln\Gamma(x)/dx$.
 The induced charge $Q_{ind}$ is negative (the function $A(Z\alpha)$ is positive,
see below.)
The distributed charge density $\rho_{distr}$ in Eq.~(\ref{rind}) is positive
and at distances $r \ll 1/M$, $\rho_{distr} \propto M^2\ln(1/Mr)$.
The density vanishes in the limit $M \to 0$, however the total distributed charge
is $-Q_{ind}$.
We comment that in conventional 3D QED, the induced charge density also consists of
  local and distributed parts.
In that case the distributed density $\rho_{distr}\propto -|e|Z\alpha/r^3$ at $r \ll 1/M$ \cite{MS}.
Interestingly,  the signs of the local and distributed charges in 3D QED
are different from those in Eq.~(\ref{rind}).
The difference is in  the leading (linear in $Z\alpha$) contribution, while the signs of
the next-to-leading contributions (non-linear vacuum polarization) are the same in both cases.
 Physically this difference is related to the fact that the effective
 charge $e^{2}$ is renormalized (increases at short distances) in conventional QED,
while in graphene the charge is not renormalized \cite{Gonzalez}.

\begin{figure}[t]
\includegraphics[width=0.48\textwidth]{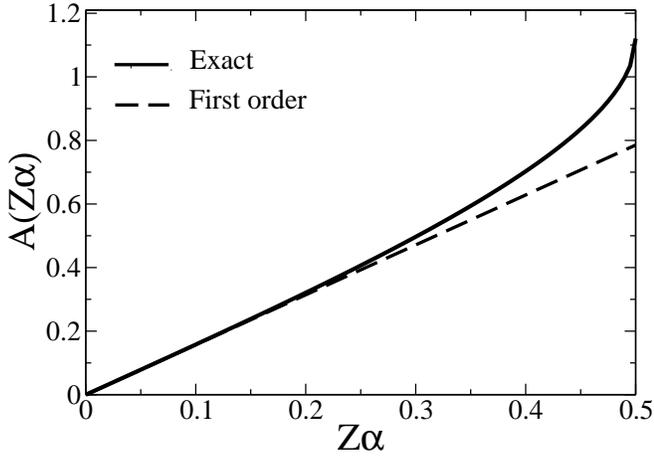}
\caption{ The induced charge  $Q_{ind}/e\equiv A$ as a function of $Z\alpha$.
 The solid line is the exact result from Eq.~(\ref{QinQRepresentation})
and the dashed line is the one loop result $A^{(1)}= \pi(Z\alpha)/2$. }
\label{fig1}
\end{figure}

The series expansion of the function $A(Z\alpha)=(\pi/2) Z\alpha + 4 \Lambda(Z\alpha)$,
 for small $Z\alpha$ reads
\begin{equation}
\label{A_series}
A(Z\alpha)= \frac{\pi}{2}(Z\alpha) +
0.783(Z\alpha)^3+1.398(Z\alpha)^5+...
\end{equation}
The first term of the expansion in Eq.~(\ref{A_series}) reflects
 the linear one-loop polarization contribution  \cite{Shytov,Subir}.
Our coefficient  of the $(Z\alpha)^3$ term  has an opposite sign with respect
 to the one found recently in  Ref.~\cite{Subir}.

The coefficients in the non-linear terms are not
small and grow as the order increases; this is in contrast with
conventional QED where they are very small and  decrease rapidly
\cite{McLerran}.
The exact function $A(Z\alpha)$ is shown in Fig.~\ref{fig1}. For
$Z\alpha>0.3$  our exact result starts to deviate substantially
from the linear order and this deviation is particularly strong as
the critical value $Z\alpha=1/2$ is approached. The exact
$A(Z\alpha)$ behavior is also stronger than the one found in a
recent numerical  calculation \cite{Shytov}, which also exhibits
lattice size dependence.
Around the critical value $Z\alpha\sim 1/2$ the function $A(Z\alpha)$ has the
following expansion
\begin{equation}
\label{A_series2}
 A(Z\alpha)= 1.12 - 1.19\sqrt{\frac{1}{2}-Z\alpha}-0.29\left(\frac{1}{2}-Z\alpha\right)+...
\end{equation}

Notice that the  induced charge $Q_{ind} = -|e| A < 0$ has a screening sign,
leading to a decrease of the effective impurity charge:
 $Z_{\mbox{eff}}=Z - A(Z\alpha)$. In fact complete screening
seems possible for $Z=1$ and $\alpha \approx 1/2$, where
$Z_{\mbox{eff}}=0$. However we will show that the self-consistent
treatment of the problem (within the  Hartree approximation) can
drastically change this  behavior.

\begin{figure}[t]
\includegraphics[width=0.48\textwidth]{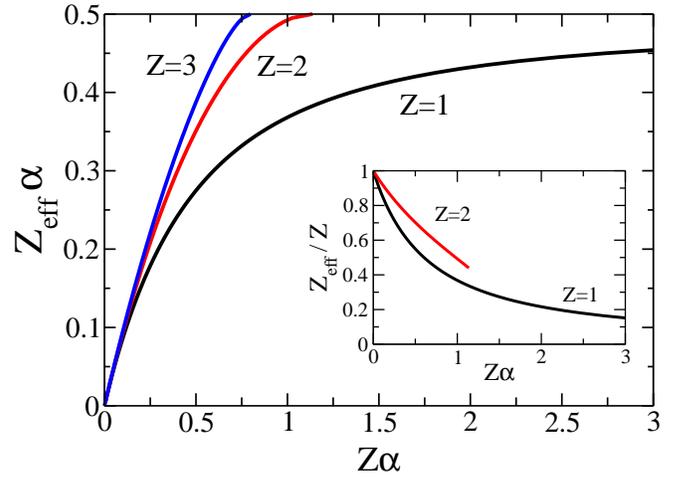}
\caption{(Color online.) The effective coupling
$Z_{\mbox{eff}}\alpha$ from Eq.~(\ref{EffCharge1}) as  a function of $Z\alpha$ for $Z=1$
(black line), $Z=2$ (red line), and $Z=3$ (blue line). Inset: Reduction of the
impurity coupling relative to its initial value, $Z_{\mbox{eff}}/Z$.}
\label{fig2}
\end{figure}

{\it Self-consistent screening}.
Since the induced charge is fully concentrated at the origin, one can easily
take into account electron-electron interactions in the Hartree approximation.
The Hartree contribution is expected to dominate over the
 Fock (exchange) contribution  for  $N \gg 1$ ($N=4$, from the spin and valley degeneracy in graphene).
To find the effective charge $Z_{\mbox{eff}}$ in the Hartree approximation, it is sufficient to
solve the following self-consistent equation
\begin{equation}
\label{EffCharge1}
Z_{\mbox{eff}}  \alpha=Z\alpha - \alpha A(Z_{\mbox{eff}}\alpha).
\end{equation}
If one uses the function $A(Z\alpha)=A^{(1)}= \pi(Z\alpha)/2$
calculated in the one-loop approximation, then  this equation is
equivalent to the usual random phase approximation (RPA).
 However, since the exact $A(Z\alpha)$  accounts for all
orders in $Z\alpha$, Eq.~(\ref{EffCharge1}) is more accurate than the RPA.

The solution $Z_{\mbox{eff}}$  of  Eq.~(\ref{EffCharge1}) is a function
of $Z$ and $\alpha$. The function $Z_{\mbox{eff}}\alpha$ versus $Z\alpha$
is shown in Fig.~\ref{fig2} for different values of the bare
charge $Z$. An impurity with charge $Z=1$ represents the most
important practical case. Interestingly, the impurity with $Z=1$
remains subcritical for all values of $\alpha$, i.e.
$Z_{\mbox{eff}}\alpha < 1/2$. An impurity with $Z=2$ becomes
critical at $\alpha_{c}^{Z=2}=0.568$, and for $Z=3$ the critical point
is $\alpha_{c}^{Z=3}=0.266$. It is clear from (\ref{EffCharge1})
that the critical $\alpha_{c}^{Z}$ for a given  bare $Z$ is described
 by the simple formula $\alpha_{c}^{Z}=0.5/(Z-A(0.5))$, where $A(0.5)=1.12$.
The fact that $A(0.5) > 1$ prevents the impurity with $Z=1$ from becoming critical.
Ultimately  the strongly non-linear variation
 of $A(Z\alpha)$ (Fig.~\ref{fig1}) is responsible for the large value  of  $A(0.5)$ and thus the
unconventional behavior  in the $Z=1$ case.

Even though the supercritical regime is
never reached for $Z=1$, the screening is very substantial, as
shown in Fig.~\ref{fig2}(Inset). For example for $\alpha = 0.5,
Z=1$, we find $Z_{\mbox{eff}}/Z = 0.55$, and similarly for $Z=2$.
Thus we find quite generally that vacuum polarization screening in
graphene is very  strong, even in the  subcritical regime.
 The  $Z=1$ case is the most relevant experimentally, since
 alkali atoms, such as potassium (K) \cite{Chen07}, typically serve
 as charged scatterers in graphene. For graphene on SiO$_2$ substrate
with typical dielectric constant $\varepsilon \approx 4$, the value of
$\alpha$ is  $\alpha \approx 0.9$
(using $e^{2} \to 2e^{2}/(1+\varepsilon)$, while the RPA correction
 is already taken into account via Eq.~(\ref{EffCharge1})). For such large $\alpha$
 the vacuum polarization effect is very strong and could be important
 for interpretation of experiments  \cite{Chen07}.

 To summarize, we have  presented an exact solution
of the Coulomb impurity screening problem in graphene in the subcritical regime.
The explicit result for the  induced vacuum polarization
 charge, Eq.~(\ref{QinQRepresentation}), is
 valid to all orders in the potential strength $Z\alpha$.
 The exact solution, when combined with the Hartree approximation,
 leads to a self-consistent solution of the impurity problem.
 An impurity with charge $Z=1$ is found to be always in the subcritical
 regime, where the induced charge is localized at the impurity site.
In this regime vacuum polarization screening is very strong
 and weakens substantially the impurity potential.
However impurities with $Z=2,3$  and higher  can become supercritical.

We are grateful to A. H. Castro Neto, V. A. Dzuba, V. M. Pereira, J. Nilsson,
and B. Uchoa  for numerous stimulating discussions.
A.I.M. gratefully
acknowledges the School of Physics at the University of New South
Wales for the warm hospitality and financial support during his visit.

\end{document}